# A Web-Based Secure Access System Using Signature Verification over Tablet PC

Fernando Alonso-Fernandez, Julian Fierrez-Aguilar, Javier Ortega-Garcia,

Joaquin Gonzalez-Rodriguez

Escuela Politecnica Superior, Universidad Autonoma de Madrid,

Ctra. Colmenar km. 15, E-28049 Madrid, Spain

Tel.: +34-91-4973363, fax: +34-91-4972235

email: Fernando.alonso@uam.es

ABSTRACT

Low cost portable devices capable of capturing signature signals are being increasingly used. Additionally, the social and legal acceptance of the written signature for authentication purposes is opening a range of new applications. In this paper, we describe a highly versatile and scalable prototype for Web-based secure access using signature verification. The proposed architecture can be easily extended to work with different kind of sensors and large scale databases. Several remarks are also given on security and privacy of network-based signature verification.

INTRODUCTION

Personal authentication in our networking society is becoming a crucial issue [1]. In this environment, there is a recent trend in using measures of physiological or behavioral traits for person authentication, which is also referred to as *biometric authentication*. Biometrics provides more security and convenience than traditional authentication methods which rely in what you know (such as a password) or what you have (such as an ID card) [2]. Within biometrics, signature verification has been an intense field of study due to its social and legal acceptance [3,4].

In this paper, we present a prototype for Web-based secure access using signature verification. The increasing use of low cost portable devices capable of capturing signature signals such as Tablet PCs, mobile telephones or PDAs is resulting in a growing demand of signature-based authentication applications. Our prototype uses a Tablet PC for signature acquisition [5] but it can be easily extended to other signature acquisition devices as well.

**WEB-BASED SECURE ACCESS USING SIGNATURE VERIFICATION**

The global architecture of our prototype is shown in Fig. 1. A *signature verification server* manages the verification process. This server communicates with a *web server*, which manages the communication with the user terminal using the HTTP protocol through a network. In our prototype, the user terminal is a Tablet PC and both the *web server* and the *signature verification server* are installed in a standard PC that communicates with the Tablet PC thorough a LAN.

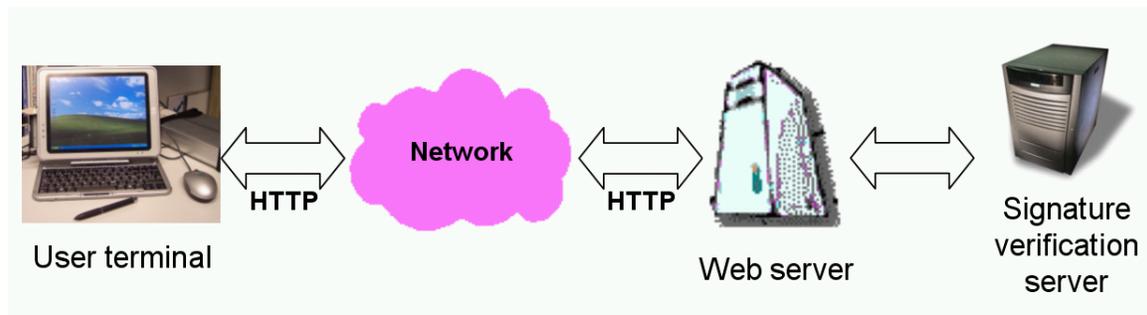

**Fig. 1: global architecture of the implemented prototype**

The proposed architecture is highly versatile. User terminal can be any device capable of capturing on-line signatures, from cheap digitizing tablets to more expensive Tablet PCs [5]. It is also highly scalable, since we can use powerful servers capable of managing several transactions in parallel, not only HTTP-based but using any other secured or unsecured protocol. Table 1 summarizes several applications that can use the proposed architecture.

This architecture can also be adapted to work in other situations such as:

- **The signature verification server has low storing capacity**. Users can be provided with a smartcard with its statistical model stored in it [6]. This approach saves considerable hard disk space in the central server and avoids the statistical models being stolen by a hacker or accidentally deleted by system administrators. On the contrary, the statistical model have to be transferred thorough a network and thus they can be intercepted by other users if no encryption or secure connection is used.

- **The signature verification server has low processing capacity**. The user terminal can then be allowed to perform the verification process, notifying to the central server the acceptation/rejection decision. This approach saves considerable processing power in the *signature verification server* and reduces the amount of data to be transferred. In addition, the user templates never are transmitted, so they cannot be intercepted. On the other hand, we need to ensure that only authorized terminals notify acceptation/rejection decisions.

Table 1: Applications of a network-based signature verification system.

| Applications | Example |
|---|---|
| e-banking | Access to bank account |
| e-commerce | Secure transactions in Internet |
| Login | Secure access to home/office computer, LAN, Web account, mobile telephone, laptop, PDA, etc. |
| POS (Point-Of-Sale) | Secure payment with credit card, verifying customers before charging their credit cards. |
| Physical access control | Secure access to restricted areas |
| Medical records management | Secure access to medical records. Only authorized users are allowed to get access. |
| e-Government | Secure operations such as ID card or driver license renovation, income tax return submission, etc. |
| Electronic data security | Access and encryption of sensitive data |

**USER ENROLMENT**

The next steps are performed in order to enroll a user in the system:

1. The user is first authorized by an administrator in the *signature verification server*. A username and a temporary password are assigned to the user. This ensures that only desired users have authorization to use the signature-based verification system.

2. Secondly, the user is requested to provide 5 signatures. These 5 signatures are used to generate a statistical model which characterizes the identity of the user [7]. The statistical model is generated in our prototype using the coordinate trajectories and pressure signals provided by the Tablet PC [5]. Technical details of the algorithm for statistical model generation can be found in [8, 9]. In our system, the user can provide its 5 signatures remotely with a downloadable application by using the temporary password assigned in the previous step. This scheme provides high flexibility. If a more secure environment is needed, another option is to enroll the users only in the presence of an administrator.

In order to account for the time variability of the signature signals, the 5 signatures used for enrolment are provided in two different sessions separated by a certain amount of time, typically 1 to 3 days. In addition, the statistical model of the user is updated along time by using the signature acquired in the last successful access.

**THE SIGNATURE VERIFICATION SERVER**

The *signature verification server* manages the verification process. It receives the requests for verification and decides if the user is accepted or not. In Fig. 2 we can see the main window of our *signature verification server*. It shows the last transactions that have been realized, which are also stored on a log file. It also allows to perform the following actions:

- User authorization, as described in the previous section.
- User management. The next information is available for each enrolled user: name, date of the last successful access, number of unsuccessful accesses since the last successful access, and block status. If a user accumulates a certain number of continuous unsuccessful accesses, he/she is blocked. In Fig. 3 we can see the user management window.
- System management. This module has the following options: storage place of the user's data, unsuccessful accesses allowed to the users, communication settings of the *signature verification server*, storage place of the log file, rules for updating the statistical models of a user, etc.

It is supposed that only authorized administrators have access to the *signature verification server*.

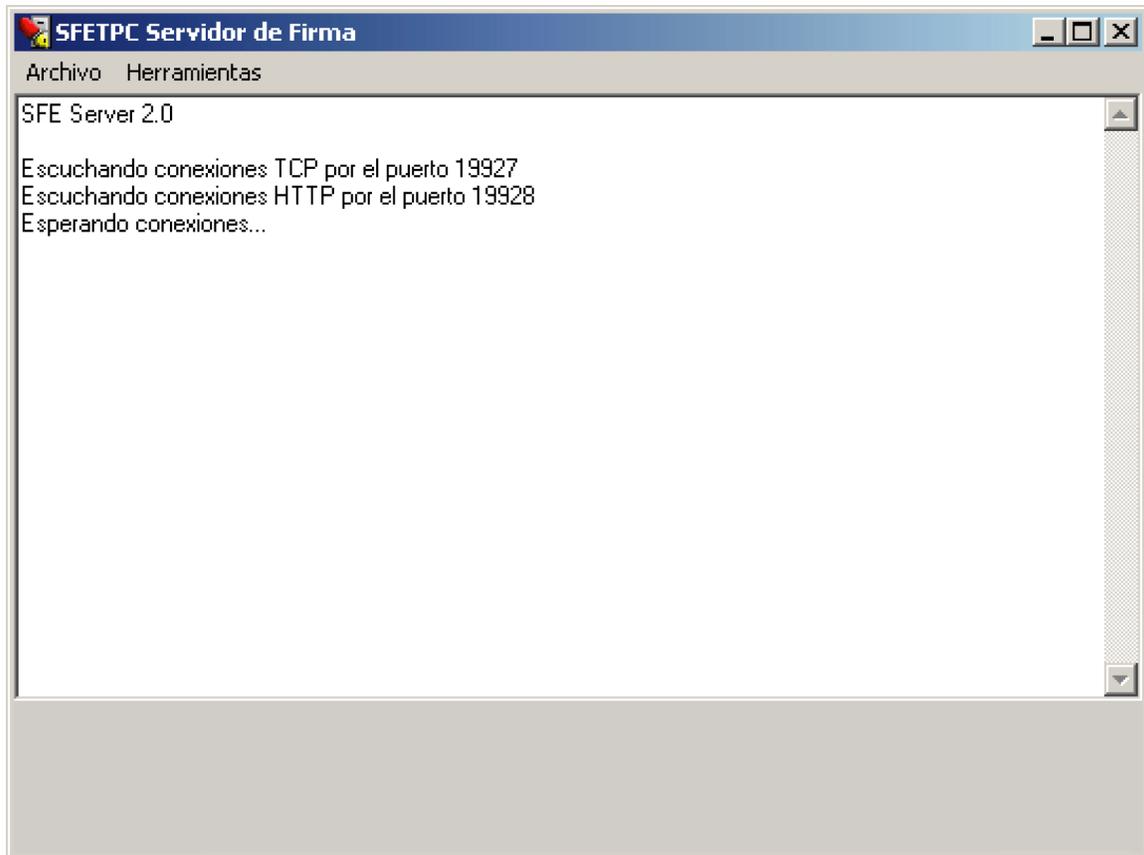

**Fig. 2: main window of the *signature verification server***

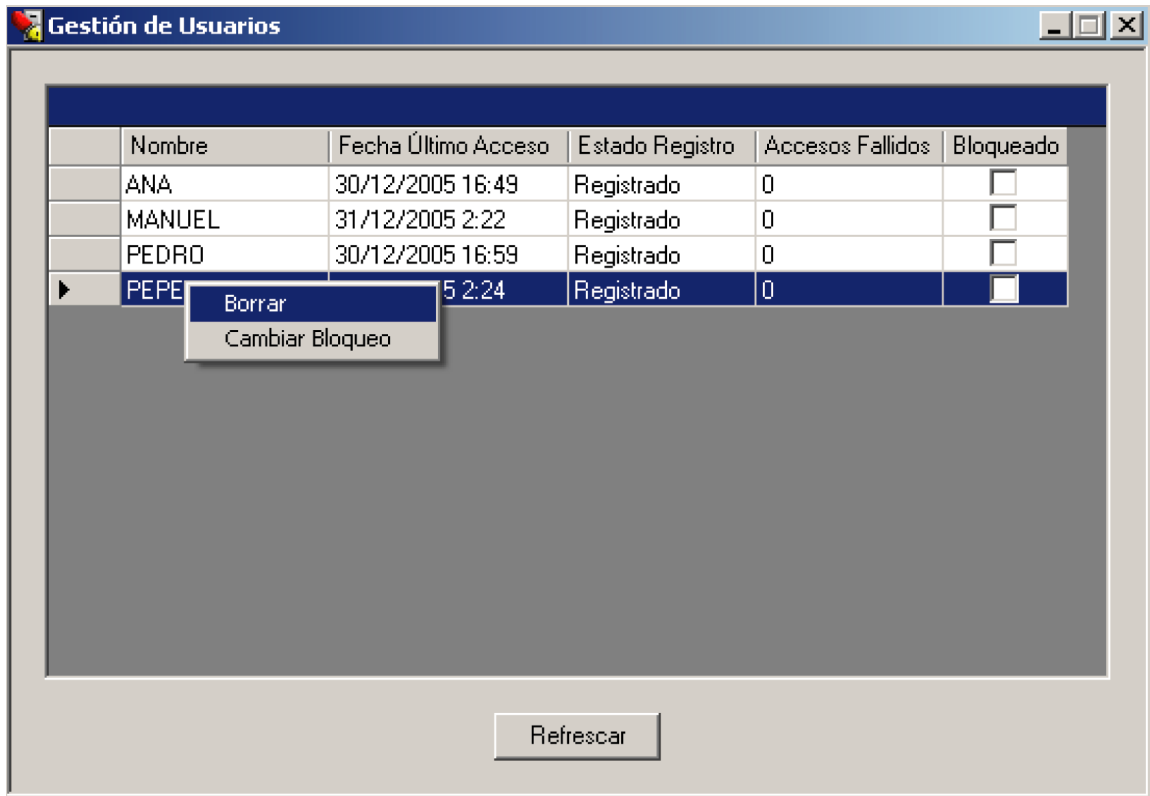

Fig. 3: user management window of the *signature verification server*

**USAGE OF THE WEB-BASED SECURE ACCESS CLIENT**

Once enrolled in the system, the user has access to the proper URL using its terminal. Fig. 4 shows the main window of our prototype, where the username and a signature realization are requested. If the user is accepted, he/she will be allowed to access to his account. If not, an appropriate message will indicate that he/she has been rejected.

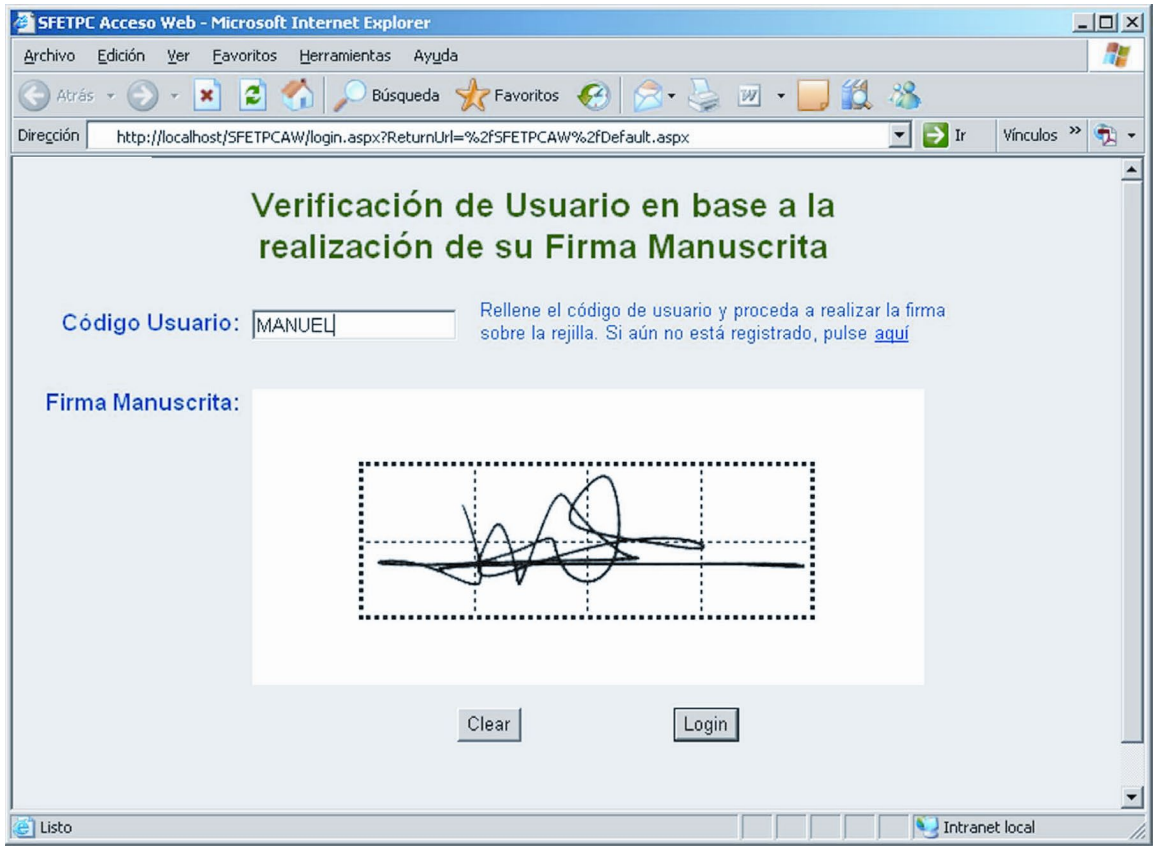

**Fig. 4: main window of our Web-based secure access prototype**

**SECURING A NETWORK-BASED SIGNATURE VERIFICATION SYSTEM**

A discussion of issues and concerns related to the design of a secure fingerprint recognition system is addressed in [10]. Some of these concerns also apply in the case of signature verification systems.

When designing a recognition system, we have to decide whether it is going to operate in verification or identification mode [1]. In verification mode, an individual who desires to be recognized claims an identity, and the system compares the captured biometric data with the biometric template corresponding to the claimed identity. In identification mode, the system recognizes an individual by comparing the captured biometric data with the templates of all the users stored in the system. If the number of users is large, verification mode is recommended unless identification is strictly necessary.

Typically, developers and integrators of systems and applications are not the producers of hardware and core software. Several factors should be taken into account when choosing hardware and software components: choose proven hardware and software technology, check standards compliance with platforms or operating systems, evaluate cost versus performance trade-off, ask for available support, etc. An SDK is usually supplied by the vendors, but system designers will usually have to develop specific applications for managing the enrolment, managing the storage and retrieval of templates and information, setting up the system options, etc.

A policy of how to deal with users with bad quality signatures has to be defined. In signature-based verification this is related to users whose signature is easy to imitate. An attended enrolment can deal with this problem, forcing users to provide signatures which are not easy to imitate, but this may result in future false rejection alarms. It is said that the security of the entire system is only as good as the weakest "password", so users with simple signatures may compromise the security of the overall application.

System administration is an important issue. The administrator may instruct users and make them familiar with the signature acquisition device. He is also in charge of the state of the acquisition devices if the verification is made in a supervised scenario. Monitoring the system log is also an important task to find out if the system is being subjected to attacks. A *threat model* for the system has to be defined and the system has to be guarded against them. The *threat model* has to be based on what needs to be protected and from whom. The typical threats in a verification system are the following:

- **Denial of Service** (DoS): the system is damaged, so legitimate users can no longer access it.
- **Circumvention**: illegitimate users gain access to the system.
- **Repudiation**: a legitimate user denies having accessed the system.
- **Covert acquisition**: trait samples of a legitimate user are obtained without his knowledge and subsequently used for illegitimate access.
- **Collusion**: illegitimate access by means of special super-users who are allowed to bypass the verification stage.
- **Coercion**: a genuine user is forced to access the system.

In Fig. 5 we can see the main modules and dataflow paths in a signature verification system. The eight possible attack points marked are: 1) Scanner, 2) Channel between the scanner and the feature extractor, 3) Feature extractor, 4) Channel between the feature extractor and the matcher, 5) Matcher, 6) Database, 7) Channel between the database and the matcher, 8) Channel between the matcher and the application requesting verification.

Note that attacks 2, 4, 7 and 8 are launched against communications channels, and are collectively called "replay" attacks. Signals in these channels can be intercepted and used at a later time. Attacks 1, 3, 5 and 6 are launched against system modules and are called Trojan horse attacks. A Trojan horse program can disguise itself as the module and bypass the true module, submitting false signals. For example, a Trojan horse program can perform a circumvention or denial-of-service (DoS) attack by always generating an acceptance or rejection decision in the matcher, respectively. Also, the sensor can be destroyed in a denial-of-service (DoS) attack.

It is very important that the feature extractor, matcher and database reside at a secure and trusted location. The scanner should implement some security capabilities (e.g.: encryption). Also, a mechanism of trust should be established between the components of the system. Mutual identification can be achieved by embedding a shared secret (e.g.: a key for a cryptographic algorithm) or by using a *Certificate Authority* (CA - an independent third party that everyone trusts and whose responsibility is to issue certificates).

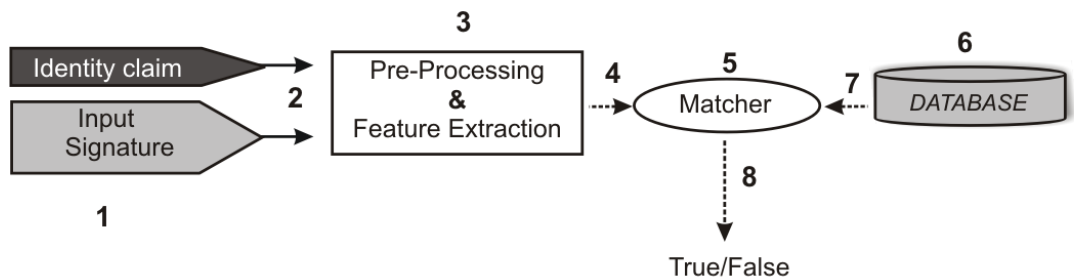

Fig. 5: design of a signature verification system. The possible security attack points are marked with numbers from 1 to 8.

**PRIVACY ISSUES**

Privacy is the ability to lead one's own life free from intrusions, to remain anonymous, and to control access to one's own personal information [2]. It is widely accepted that biometric identifiers provide positive person recognition better than conventional technologies (token-based or knowledge-based). But several arguments and objections are given against biometric recognition: hygiene of biometric scanners that require contact; negative connotations associated with some biometrics used in criminal investigation (DNA, fingerprint, face); inference of information from biological measurements; linkage of biometric information between different applications, allowing to track individuals, either with or without permission; acquisition of biometric samples without knowledge of the person, allowing covert recognition of people; etc. The abuse of biometric information is an open issue that should be addressed by governments, industry and organizations. Unless a consensus is reached, citizens may be reluctant to provide biometric measurements and to use biometric recognition systems.

One way to deal with some of the associated privacy problems is the use of systems with the information in a decentralized place over which the individual has complete control. For example, a smartcard can be issued with the template of the user stored in it [6]. Even more, as the computational power of smartcards is continuously increasing, it will be possible to implement the verification step inside the card in a match-on-card architecture. The card will only have to deliver the acceptation/rejection decision. In that case, neither the template of the user nor the acquired biometric samples are sent to any centralized application.

**CONCLUSIONS**

A prototype for Web-based secure access using signature verification has been described. The proposed architecture ensures high versatility and scalability. The *signature verification server,* which manages the verification process, is capable of communicating with a variety of sensors through several kinds of networks using standard protocols such as HTTP. It can be customized depending on factors such as: allowed number of users, cost of the acquisition sensors, network used in the access, storing or processing capacity of the *signature verification server*, etc.

Several issues have to be taken into account when designing a network-based signature verification system: mode of operation (verification or identification), selection of hardware and software components, policy with users with bad quality signatures, administration of the system, definition of a *threat model*, detection of attacks and implementation of a mechanism of trust between components of the system. Privacy issues have to be also considered when designing a system based on biometric information.

ACKNOWLEDGEMENTS

This work has been supported by BBVA, BioSecure NoE and the TIC2003-08382-C05-01 project of the Spanish Ministry of Science and Technology. F. A.-F. and J. F.-A. thank Consejeria de Educacion de la Comunidad de Madrid and Fondo Social Europeo for supporting their PhD studies.

FERNANDO ALONSO-FERNANDEZ received the M.S. degree in Electrical Engineering in 2003, from Universidad Politecnica de Madrid. Since 2004 he is with the Computer Engineering Department, Universidad Autonoma de Madrid, where he is currently working towards the Ph.D. degree on biometrics. He has published over 10 international contributions. His research interests include signal and image processing, pattern recognition and biometrics.

JULIAN FIERREZ-AGUILAR received the M.S. degree in Electrical Engineering in 2001, from Universidad Politecnica de Madrid. Since 2004 he is with the Computer Engineering Department, Universidad Autonoma de Madrid, where he is currently working towards the Ph.D. degree on multimodal biometrics. His research interests include signal and image processing, pattern recognition and biometrics. He has published over 35 international contributions. He was the recipient of the Best Poster Award at AVBPA 2003, the Rosina Ribalta Award from the Epson Iberica Foundation to the best PhD Proposal in 2005, the Best Student Paper at ICB 2006, and led the development of the UPM signature verification system ranked 2nd in SVC 2004.

JAVIER ORTEGA-GARCIA received the Ph.D. degree in electrical engineering in 1996 from Universidad Politecnica de Madrid. He is currently an Associate Professor at the Computer Engineering Department, Universidad Autonoma de Madrid. His research interests are focused on forensic acoustics and biometrics signal processing. He has published over 50 international contributions. He has participated in several scientific and technical committees, and has chaired "Odyssey-04, The ISCA Speaker Recognition Workshop".

JOAQUIN GONZALEZ-RODRIGUEZ received the Ph.D. degree in electrical engineering in 1999 from Universidad Politecnica de Madrid. He is currently and Associate Professor at the Computer Engineering Department, Universidad Autonoma de Madrid. His research interests are focused on signal processing, biometrics and forensics. He is an invited member of ENFSI and has been vice-chairman for "Odyssey-04, The ISCA Speaker Recognition Workshop".